# Graph invariants for unique localizability in cooperative localization of wireless sensor networks: rigidity index and redundancy index

Tolga Eren
Department of Electrical and Electronics Engineering, Kirikkale University
Kirikkale, 71450 Turkey
Correspondence: erente@gmail.com
Web: http://tolgaeren.blogspot.com

**Abstract**

*Rigidity theory enables us to specify the conditions of unique localizability in the cooperative localization problem of wireless sensor networks. This paper presents a combinatorial rigidity approach to measure (i) generic rigidity and (ii) generalized redundant rigidity properties of graph structures through graph invariants for the localization problem in wireless sensor networks. We define the rigidity index as a graph invariant based on independent set of edges in the rigidity matroid. It has a value between 0 and 1, and it indicates how close we are to rigidity. Redundant rigidity is required for global rigidity, which is associated with unique realization of graphs. Moreover, redundant rigidity also provides rigidity robustness in networked systems against structural changes, such as link losses. Here, we give a broader definition of redundant edge that we call the "generalized redundant edge." This definition of redundancy is valid for both rigid and non-rigid graphs. Next, we define the redundancy index as a graph invariant based on generalized redundant edges in the rigidity matroid. It also has a value between 0 and 1, and it indicates the percentage of redundancy in a graph. These two indices allow us to explore the transition from non-rigidity to rigidity and the transition from rigidity to redundant rigidity. Examples on graphs are provided to demonstrate this approach. From a sensor network point of view, these two indices enable us to evaluate the effects of sensing radii of sensors on the rigidity properties of networks, which in turn, allow us to examine the localizability of sensor networks. We evaluate the required changes in sensing radii for localizability by means of the rigidity index and the redundancy index using random geometric graphs in simulations.*

## 1. Introduction

Localization is an essential service for many applications of wireless sensor networks. A wireless sensor network consists of a small number of *anchors* (reference nodes) and a large number of small, cheap *ordinary nodes* (non-anchors). Anchors have a priori knowledge of their own positions, e.g., GPS, and ordinary nodes have no prior knowledge of their locations. If ordinary nodes were capable making measurements to multiple anchors, they could determine their positions. However, several ordinary nodes cannot directly communicate with anchors because of power limitations or signal blockage. A recent paradigm is cooperative localization, in which ordinary nodes help each other to determine their locations [1,2]. In cooperative localization, ordinary nodes not only make measurements with anchors, but also they make measurements with other ordinary nodes. The types of measurements usually include distance estimates and/or angle estimates [3].

The set of ordinary nodes is uniquely localizable if there is a unique set of positions satisfying the conditions resulting from measurements. Unique solvability of the cooperative localization problem is characterized by the results from "rigidity theory."

Redundant rigidity is required for global rigidity, which is associated with unique realization of graphs. Use of rigidity theory in localization is well described in the literature [4−10]. More details will be given in Section 2.

Recent works introducing the rigidity theory into formation control has also provided provably correct methods to model and analyze the ad hoc network topologies within robotic teams [11-15]. For example, rigidity theory provides us tools for formation control using relative distance measurements instead of relative position measurements. Moreover, rigidity is necessary to estimate relative positions using only relative distance measurements [11].

Quantitative measures of rigidity have been proposed by researchers recently. Jacobs et al. provided a measure of rigidity within the context of microstructures of proteins, and their approach are based on chemical bonds [16]. Zhu and Hu studied quantitative measure of formation rigidity using stiffness matrix [17]. Zelazo et al. introduced the rigidity eigenvalue based on symmetric rigidity matrix [11]. The latter two studies employed rigidity based matrices and studied the properties of those matrices to explore the rigidity properties of networks.

In this paper, first we provide a measure of "generic rigidity" in 2-space for both rigid and non-rigid graphs. Then the concept of generalized redundancy is introduced, which allows us to provide a measure of "generalized redundancy" in 2-space for both rigid and non-rigid graphs. Our approach is based on the combinatorial characterizations of rigidity and redundancy. Specifically, the main contributions of this work are: (a) the translation of edge distribution in a network graph to that of a rigidity measure that we term the "rigidity index," (b) the translation of the generalized redundancy of edge distribution in a network graph to that of a redundancy measure on network rigidity that we term the "redundancy index."

From a graph theory point of view, the benefits of these measures are as follows: They permit us (i) to quantify the distribution of edges in a graph in terms of rigidity and redundant rigidity, (ii) to compare various graphs for rigidity in terms of the distribution of their redundant and non-redundant edges.

From a sensor network point of view, these two measures enable us to evaluate the effects of sensing radii of sensors on the rigidity and redundancy properties of networks, which in turn allows us to examine the localizability of sensor network graphs. In particular, we are interested in the following questions: (i) how much change in sensing radii do we need to reach from non-rigidity to rigidity, and to reach from rigidity to redundant rigidity in random geometric graphs? (ii) given that redundant rigidity is associated with unique localizability, is redundant rigidity a heavy burden on the network once rigidity is achieved? We provide answers to these questions in this paper.

The localization process often needs to be repeated in mobile wireless sensor networks. Mobility brings the possibility of the loss of links, which enforces to have not only localizable network structures but also structures which remain localizable after the loss of links in the network. Since redundancy plays a role in robustness, redundancy measure also helps us to evaluate robustness to link losses.

The structure of the paper is as follows. We give preliminaries on rigidity in Section 2. Main results on the rigidity index, the redundancy index and the corresponding complexity analysis are provided in Section 3. Examples to illustrate those indices on graphs are presented in Section 4. Applications of these two indices in sensor network simulations are demonstrated in Section 5. Finally, the paper ends with a conclusion and some outlook on future directions in Section 6.

## 2. Rigidity

First we provide below some background on rigidity, redundant rigidity and global rigidity. We refer the reader to [18-21] and the references therein for more details.

### 2.1 Rigid frameworks and the rigidity matrix

We model a network by a finite graph $G = (V, E)$. All graphs considered are finite without loops and multiple edges. Nodes of the network correspond to the vertices of $G$, and for every link in the network there is an edge joining the corresponding vertices of the graph. A *framework* $G(p)$ is a graph $G = (V, E)$ and a plane configuration $p : V \to R^2$. Two frameworks $G(p)$ and $G(q)$ are *equivalent* if $\|p(v_i) - p(v_j)\| = \|q(v_i) - q(v_j)\|$ holds whenever $v_iv_j$ corresponds to an edge of $G$, where $\|.\|$ denotes the distance. $G(p)$ and $G(q)$ are *congruent* if for any two vertices $v_i$, $v_j \in V$, $\|p(v_i)-p(v_j)\| = \|q(v_i)-q(v_j)\|$ holds. A framework $G(p)$ in $R^2$ is *rigid* if there is an $\varepsilon > 0$ such that for any other configuration $q$ in $R^2$, where $\|p(v) - q(v)\| < \varepsilon$ for all $v$ in $V$ and $G(p)$ is equivalent to $G(q)$, then $p$ is congruent to $q$. Intuitively, we may consider the rigidity of bar-joint frameworks. Here, bars correspond to edges, and joints correspond to vertices. A bar-joint framework is rigid if it has only trivial deformations, e.g., translations and rotations.

The *rigidity matrix* $R(G,p)$ of a framework $G(p)$ is the $|E| \times 2|V|$ matrix, whose rows correspond to the edges and whose columns correspond to the coordinates of the vertices, where $|.|$ denotes the cardinality of a set. If $e = v_iv_j \in E$, then the entry in the row $e$ and the column $v_i$ is $p(v_i) - p(v_j)$, the entry in the row $e$ and the column $v_j$ is $p(v_j) - p(v_i)$, and the other entries in the row $e$ are zeros. If $e = v_iv_j$ is not in $E$, then the entire row $e$ is zeros. A framework $(G, p)$ is called *infinitesimally rigid* if rank$\{R(G, p)\} = 2|V| - 3$. Infinitesimal rigidity of $(G, p)$ implies rigidity. The converse is not true in general. However, if $p$ is generic a configuration then rigidity also implies infinitesimal rigidity. Configuration $p$ is *generic* if the coordinates of the points do not satisfy any non-zero polynomial equation with integer coefficients.

### 2.2 Rigid graphs

A graph $G$ is *rigid* in $R^2$ if $(G, p)$ is rigid for every generic configuration $p$. A graph $G = (V, E)$ is called *minimally rigid* if $G$ is rigid, and $G$ loses its rigidity property when any one of the edges in $E$ is removed. The combinatorial characterization of rigidity in 2-space was first given by Laman in [22]. For a subset $V' \subseteq V$ let $G' = (V', E')$ denote the subgraph of $G$ induced by $V'$.

**Theorem (Laman** [22]**)** $G = (V, E)$ is minimally rigid if and only if $|E| = 2|V| - 3$ and

$$|E'| \leq 2 |V'| - 3 \text{ for all } V' \subseteq V \text{ with } |V'| \geq 2. \quad (1)$$

### 2.3 Global rigidity and redundant rigidity

A framework $(G, p)$ is *globally rigid* if, for every configuration $q$, framework $(G, q)$ which is equivalent to framework $(G, p)$ is congruent to $(G, p)$. The essential characteristic of global rigidity is that the distance between every pair of nodes is preserved for different framework realizations, and not just those defined by the edge set. If a graph $G = (V, E)$ is rigid but not minimally rigid, then $G$ is called a *redundantly rigid* graph. For such a graph, $G - e$ is rigid for all $e \in E$. An edge is called a *redundant edge* if graph remains rigid after its removal. It is known that $G$ has a unique generic realization in 2-space if and only if $G$ is 3-connected and redundantly rigid [20]. A graph satisfying these two conditions, i.e., 3-connectivity and redundant rigidity, is called a globally rigid graph.

## 3. Measures of Generic Rigidity and Generalized Redundant Rigidity

### 3.1 Rigidity index

Let $E$ be a finite set and let $I$ be a family of subsets of $E$. Then $I$ forms the independent sets of a matroid $M(G)$ if it satisfies the following three matroid axioms [20]:

(A1) $\varnothing \in I$,

(A2) if $E' \in I$ and $E'' \subseteq E'$ then $E'' \in I$,

(A3) if $E_1$ and $E_2$ are in $I$, and $|E_1| < |E_2|$, then there is an element $e$ in $E_2 - E_1$ such that $E_1 \cup e \in I$.

We refer the reader to the book of Oxley [23] for more information on matroid theory.

The rigidity matroid of the framework $(G, p)$ is defined by linear independence of rows of the rigidity matrix $R(G,p)$. For a graph $G = (V, E)$, let $E'$ be a non-empty subset of $E$, $V'$ be the set of vertices incident with $E'$. Then $E'$ is called *independent* if

$$|E'| \leq 2 |V'| - 3 \text{ for all } E' \subseteq E. \quad (2)$$

We note that Laman's Theorem characterizes the bases of the rigidity matroid of the complete graph on $V$ [21].

Let $G = (V, E)$ be a graph. Recall that a set of edges $E' \subseteq E$ is independent if the subgraph induced by $E'$ satisfies Equation (2). Let $S$ be the collection of the sets satisfying Equation (2). We define the *rigidity index*, denoted by $K_r(G)$, as follows:

$$K_r(G) := \frac{\max_{E' \in S} |E'|}{2|V| - 3}. \tag{3}$$

This index is the ratio of the size of the set of independent edges over the maximal number of independent edges. Note that $0 \leq K_r(G) \leq 1$. A value of zero indicates the empty set, i.e., no links in the network. A larger value of $K_r$ indicates that the network is closer to rigidity. A value of 1 indicates a rigid network.

**3.2 Redundancy index**

Recall that an edge $e$ in a rigid graph is redundant if the graph remains rigid after the removal of $e$. Here we give a broader definition of redundant edge that we call "generalized redundant edge." This definition of redundancy is valid for both rigid and non-rigid graphs. Let $G = (V, E)$ be a graph, where $G$ is not necessarily rigid. An edge $e \in E$ is called a *generalized redundant edge* if

$$K_r(G - e) = K_r(G).$$

Let the set of generalized redundant edges in graph $G = (V, E)$ be denoted by $E_u(G)$. Then the *redundancy index*, denoted by $K_u(G)$, is defined as,

$$K_u(G) := \frac{|E_u(G)|}{|E|}.$$

Note that $0 \leq K_u(G) \leq 1$. A value of zero indicates that there are no redundant edges in $E$. On the other hand, if $K_u(G) = 1$, then this means that each edge in $E$ is redundant. A larger value of $K_u$ indicates that a network is more robust to main its rigidity index.

It may suffice to know the redundancy index for most of the applications. However, it is possible to define higher types of redundancy indices, and that is what we will discuss next.

Let $G = (V, E)$ be a graph. A pair of edges $(e_i, e_j)$, where $e_i, e_j \in E$ is a *generalized redundant pair* if

$$K_r(G - \{e_i, e_j\}) = K_r(G).$$

Let the set of "all" generalized redundant pairs be denoted by $E_u^2$. Let $E^2$ denote the set of "all pairs" of edges in $E$. Note that $|E^2| = \binom{|E|}{2}$. Let $K_u^2(G)$ denote the *redundancy index of type 2*. Then $K_u^2(G)$ is defined by

$$K_u^2(G) := \frac{|E_u^2|}{|E^2|}.$$

We generalize the notion of redundancy index for the removal of $k$ edges, where $k \leq |E| - \max_{F \in S} |F|$ and $S$ is the collection of the sets satisfying Equation (2). A *k-tuple* of edges $(e_1, e_2, \ldots, e_k)$ is a *generalized redundant k-tuple* if

$$K_r(G - \{e_1, e_2, e_3, \ldots, e_k\}) = K_r(G).$$

Let the set of "all" redundant *k-tuples* be denoted by $E_u^k$. Let $E^k$ denote the set of "all *k-tuples*" of edges in $E$. Note that $|E^k| = \binom{|E|}{k}$. Let $K_u^k(G)$ denote the *redundancy index of type k*. Then $K_u^k(G)$ is defined by

$$K_u^k(G) := \frac{|E_u^k|}{|E^k|}. \qquad (4)$$

Note that the redundancy index $K_u(G)$ defined above may also be called the "redundancy index of type 1," and may be denoted as $K_u^1(G)$ to be consistent with the higher types of redundancy indices. However, in this paper, we will refer it as the "redundancy index" and use the notation $K_u(G)$ for convenience.

### 3.3 Complexity analysis of computing the rigidity index and the redundancy index

For a graph $G$ with $n$ vertices, there are polynomial algorithms for testing rigidity in $O(n^2)$ time. Performing the independence test to compute the rigidity index takes the total running time of $O(n^2)$ [21].

It is possible to test redundant rigidity, by removing one edge at a time, and by using the algorithm for rigidity repeatedly. So determining the redundancy index has $O(mn^2)$ complexity, since rigidity has to be tested for the removal of each edge, where $m$ is the number of edges in $G$. In general, determining the redundancy index of type $k$, where $k \leq |E| - \max_{F \in S}|F|$, has $O(m^k n^2)$ complexity, since rigidity has to be tested for the removal of every $k$-combination of $m$ edges. There are more efficient algorithms for testing redundant rigidity in the literature, where redundant rigidity in 2-space can be decided in $O(n^2)$ time. See [21,24] for more details on the algorithmic aspects.

## 4. Examples

In this section, we provide illustrative examples to demonstrate the rigidity index and the redundancy index.

A minimally rigid graph $G_1(V, E_1)$ is shown in Figure 1a. There are $2|V| - 3$ edges in $G_1$ and the generic rigidity index $K_r(G_1)$ is 1. All the edges in $E_1$ are independent since there is no redundancy. Therefore the redundancy index $K_u(G_1)$ is zero. In fact, all the redundancy indices of higher orders are also zero.

A non-minimal rigid graph $G_2(V, E_2)$ is shown in Figure 1b. Its rigidity index $K_r(G_2)$ is 1. Note that $|E|=18$ and $E_u(G_2) = \{(v_1,v_2), (v_1,v_3), (v_1,v_4), (v_1,v_5), (v_2,v_3), (v_2,v_4), (v_2,v_5), (v_3,v_4), (v_3,v_5), (v_4,v_5)\}$. Therefore, its redundancy index $K_u(G_2) = |E_u(G_2)| / |E| = 10/18 \approx 0.5556$.

In Figure 2a, $G_3(V, E_3)$ is an example for a non-rigid graph with no redundant edges. All the edges are independent since there is no redundancy. Therefore $\max_{F \in S}|F| = 13$. Since the number of vertices is $|V| = 9$, maximal number of independent edges is 2|V| − 3 = 15. Therefore, the rigidity index $K_r(G_3)$ = 13/15 ≈ 0.8667. Since there is no redundancy at all in this graph, its redundancy index $K_u(G_3)$ is zero.

In Figure 2b, $G_4$ (V, $E_4$) is a non-rigid graph with redundant edges. The number of independent edges is 14. Since the number of vertices is |V| = 9, 2|V| − 3=15. Therefore its rigidity index $K_r(G_4)$ = 14/15 ≈ 0.9333. The set of generalized redundant edges is $E_u(G_4)$ = {($v_1$,$v_2$), ($v_1$,$v_3$), ($v_1$,$v_6$), ($v_2$,$v_3$), ($v_2$,$v_6$), ($v_2$,$v_7$), ($v_3$,$v_6$), ($v_3$,$v_7$), ($v_6$,$v_7$)}. The total number of edges is 16 in this graph. So its redundancy index $K_u(G_4)$ = 9/16 = 0.5625.

## 5. Simulations

### 5.1 Simulation using random geometric graphs

In our simulations, we use random geometric graphs in modeling sensor networks. A set of *n* vertices are placed independently and uniformly at random. If the distance between any two vertices is at most the sensing radius $r_s$, then there is an edge between these two vertices [4].

A uniform random distribution of *n* sensor nodes (*n* = 25) in an area of *d×d*, where the side length of the area *d* = 30 units, is shown in Figure 3. A depiction of a resulting graph for $r_s$ = 12 units, i.e., 40% of *d* (the side length of the area), is shown in Figure 4. For this random geometric graph, $K_r(G)$ = 1 and $K_u(G)$ = 1.

As $r_s$ changes, the values of the rigidity index $K_r(G)$, and the redundancy index $K_u(G)$ do also change. Plot of the rigidity index and the redundancy index against the ratio $r_s / d$ between the sensing radius $r_s$ and the side length of the area *d*, for the node distribution in Figure 3, are shown in Figure 5.

### 5.2 Average results of simulations

We generated 50 different uniform random distributions of 25 sensor nodes in an area of 30×30. Then, for each node distribution, we computed $K_r(G)$ and $K_u(G)$ as a function of the ratio $r_s/d$ between the sensing radius $r_s$ and the side length of the area *d*. Next, we took the average of these fifty simulations. The resulting plot is shown in Figure 6. Figure 7 presents a zoomed-in version of Figure 6 focusing on the region where $K_r(G)$ and $K_u(G)$ are getting close to the value of 1. For the average result, $K_r(G)$ becomes 1 when $r_s/d$ = 0.49, and $K_u(G)$ becomes 1 when $r_s/d$ = 0.57.

### 5.3 Discussion

From the simulations of the average result, we see that a network graph reaches rigidity when $r_s/d$ is approximately 0.49, and it reaches redundant rigidity when $r_s/d$ is approximately 0.57. Therefore, from the average result of $K_r(G)$ and $K_u(G)$, we observe that approximately 16 percent increase in $r_s$ transforms a rigid graph into a redundantly rigid graph. We conclude that achieving redundant rigidity imposes considerably less burden on the network once rigidity is achieved. Recall that these were the issues that we posed at the beginning of this paper. We also note that, for both the individual and the average result, $K_r(G)$ is a non-decreasing function of the ratio $r_s/d$, while $K_u(G)$ is a non-monotonic function.

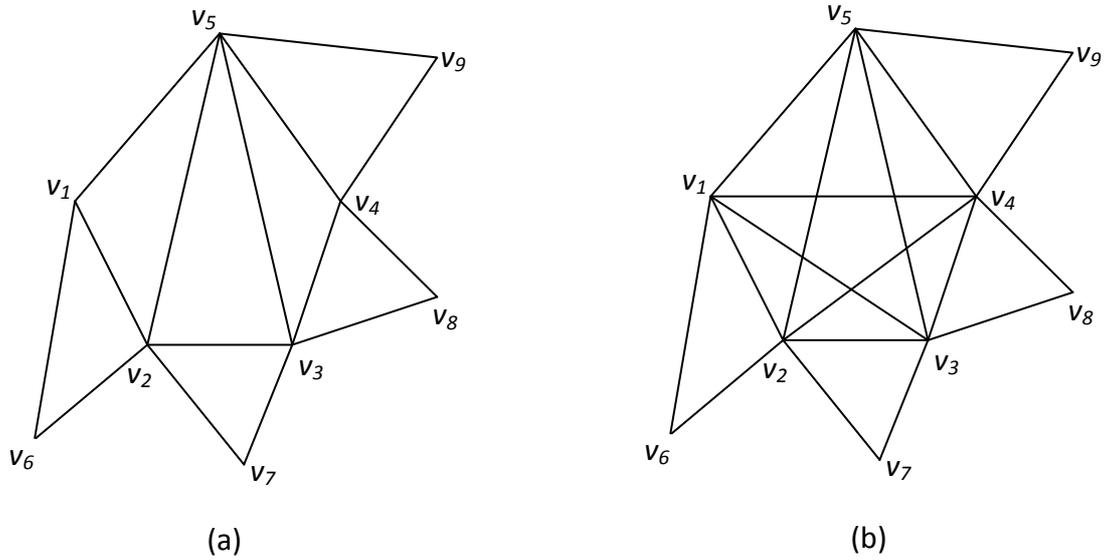

**Figure 1.** (a) $G_1(V, E_1)$ is an example for a minimally rigid graph, and its rigidity index $K_r(G_1) = 1$, and its redundancy index $K_u(G_1)$ is zero. (b) $G_2(V, E_2)$ is a non-minimal rigid graph, and its rigidity index $K_r(G_2) = 1$, and its redundancy index, $K_u(G_2) = 10/18 \approx 0.5556$.

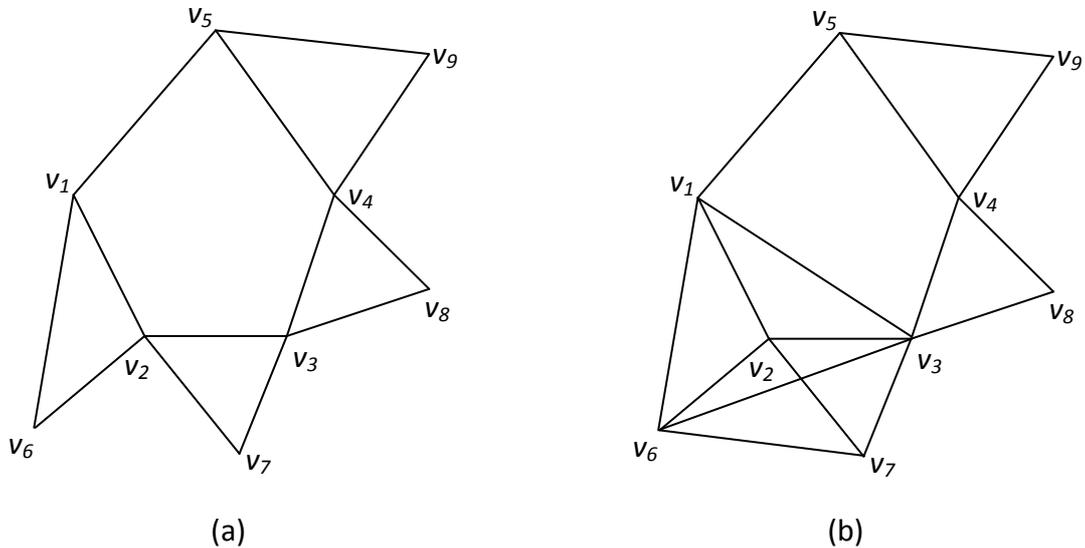

**Figure 2.** (a) $G_3(V, E_3)$ is an example for a non-rigid graph with no redundant edges, and its rigidity index $K_r(G_3) = 13/15 \approx 0.8667$, and its redundancy index $K_u(G_3)$ is zero. (b) $G_4(V, E_4)$ is a non-rigid graph with redundant edges, and its rigidity index $K_r(G_4) = 14/15 \approx 0.9333$, and its redundancy index $K_u(G_4) = 9/16 = 0.5625$.

## 6. Conclusion and Future Work

This work presented identifications of graph invariants for generic rigidity and redundant rigidity that we termed the rigidity index and the redundancy index. These indices are scalars resulting from the combinatorial rigidity properties of a network graph. The results were demonstrated with examples and simulations.

We explored the transition from non-rigidity to rigidity and from rigidity to redundant rigidity. These indices are quantitative measures of network rigidity and they have properties that meet the demands in applications. Specifically, these indices are helpful to determine the necessary increase in sensing radii to transform a non-rigid graph to a rigid graph, and transform a rigid graph to a redundantly rigid graph.

Redundancy index can also be considered as a measure of robustness. A larger redundancy index indicates a more robust network against link losses. From a network point of view, redundant rigidity in networked systems is essential in unknown environments for a reliable performance against structural changes, where redundant rigidity ensures to maintain rigidity properties of the network when some communication links may become unavailable because of the dynamic conditions of the changing environment, such as obstruction.

Higher types of redundancy indices as defined in Section 3 will provide us information on robustness to multiple link losses and these indices deserve further investigation, which also constitutes our future work. Another possible future research direction is to determine and examine the rigidity index and the redundancy index in 3-dimensional space.

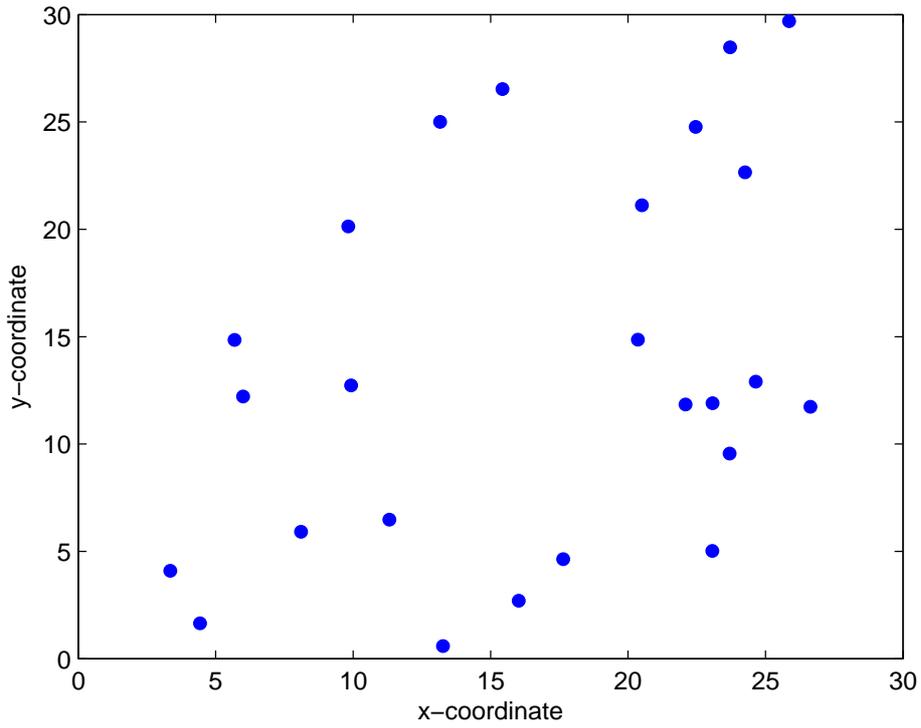

**Figure 3.** A uniform random distribution of *n* sensor nodes (*n* = 25) in an area of *d* × *d*, where *d* = 30 units.

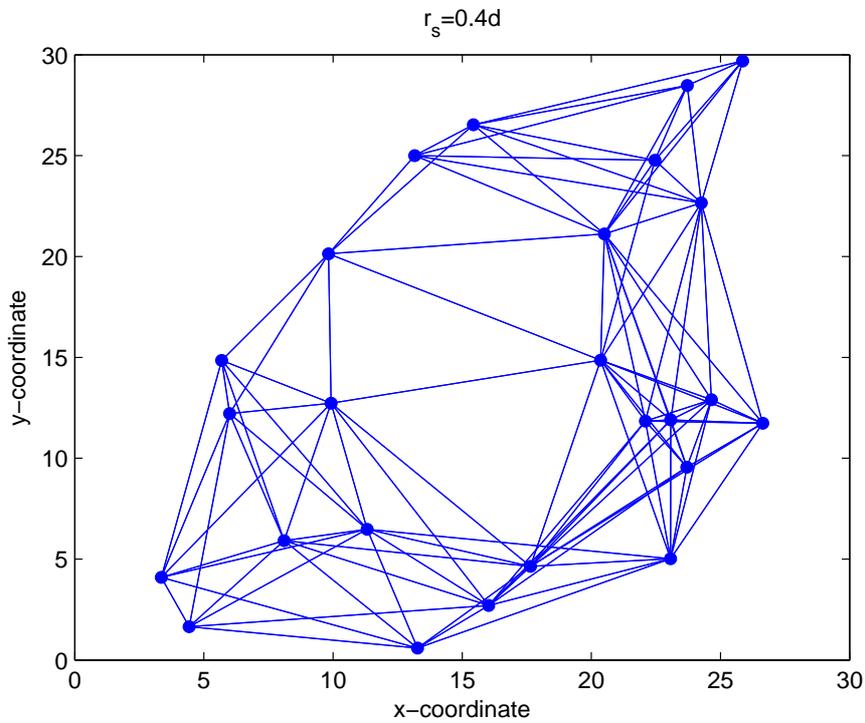

**Figure 4.** The random geometric graph resulting from the node distribution shown in Figure 3. Sensing radius $r_s$ =12 units, i.e., 40% of *d* (the side length of the area) for this particular simulation.

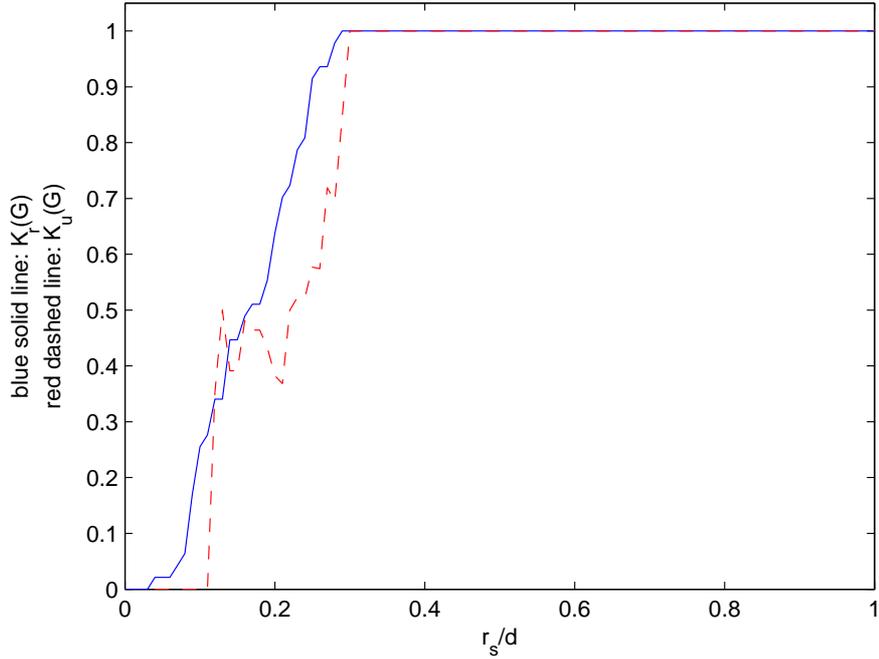

**Figure 5.** Plot of the generic rigidity index $K_r(G)$ (plotted with solid blue line)**,** and the redundancy index $K_u(G)$ (plotted with red dashed line), against the ratio $r_s/d$ between the sensing radius $r_s$ and the side length of the area $d$, for the node distribution shown in Figure 3.

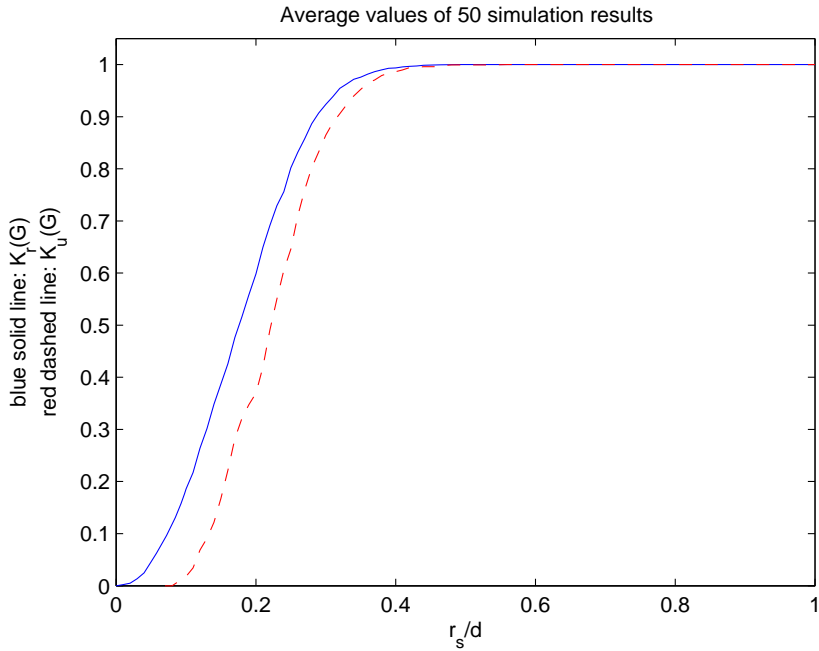

**Figure 6.** Plot of the average rigidity index $K_r(G)$**,** and the average redundancy index $K_u(G)$, against the ratio $r_s/d$ between the sensing radius $r_s$ and the side length of the area $d$. A total of fifty different uniform node distributions are used in calculating the average.

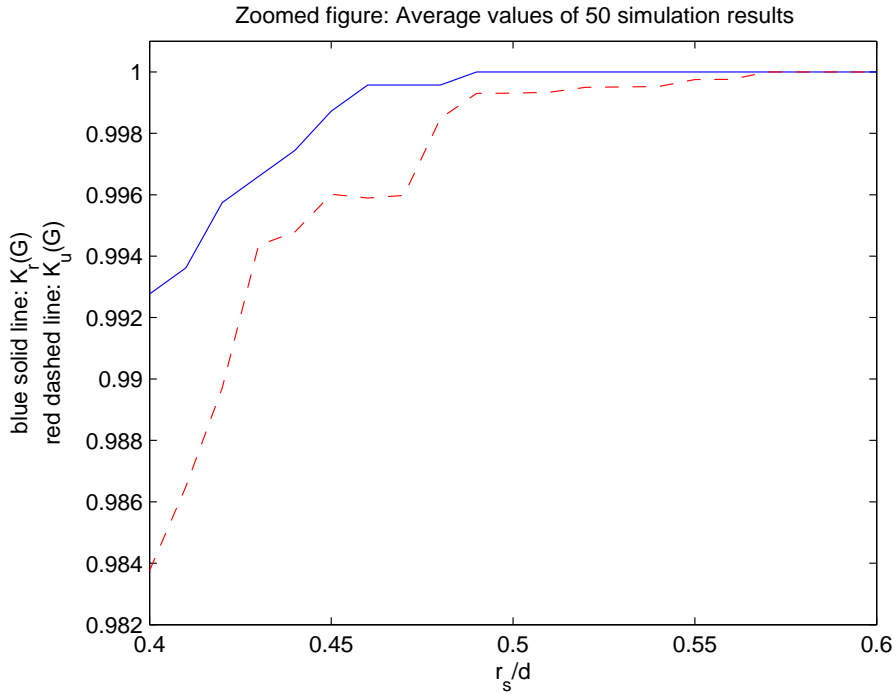

**Figure 7.** Zoomed-in version of Figure 6 focusing on the region where $K_r(G)$ and $K_u(G)$ are getting close to the value of 1.